\documentclass[twocolumn,prl,showpacs,floatfix]{revtex4}

\usepackage{graphicx}
\usepackage{dcolumn}
\usepackage{bm}
\usepackage{amsmath}

\begin{document}

\title{
Crystalline character of high-magnetic-field cusp states in quantum dots
}

\author{Constantine Yannouleas}
\author{Uzi Landman}

\affiliation{School of Physics, Georgia Institute of Technology,
             Atlanta, Georgia 30332-0430}

\date{31 December 2003; Revised 1 March 2004}

\begin{abstract}
Conditional probability distributions from exact diagonalization are used to
investigate the crystalline or liquid character of the downward cusp states in
parabolic quantum dots (QD's) at high magnetic fields. These states are 
crystalline in character for fractional fillings covering both low and high 
values, unlike the liquid Jastrow-Laughlin wave functions, but in remarkable 
agreement with the rotating-Wigner-molecule ones [Phys. Rev. B {\bf 66}, 
115315 (2002)]. The cusp states are precursors to the bulk fractional 
quantum Hall states (and not to the bulk Wigner crystal), since the collective 
rotation stabilizes the {\it rotating\/} Wigner molecule (formed in the
QD) relative to the {\it static\/} one.
\end{abstract}


\maketitle
The excitation energy spectra of two-dimensional $N$-electron semiconductor 
quantum dots (QD's) in high magnetic fields $(B)$ exhibit downward cusps  
\cite{lau1,mak1,jai1,yl1} at certain magic angular momenta, corresponding to
states with enhanced stability. These cusp states have been long recognized 
\cite{yang,jai1,haw,yl1} as the finite-size precursors of
fractional quantum Hall states in the bulk. In the literature of the 
fractional quantum Hall effect (FQHE), ever since the celebrated paper
\cite{lau2} by Laughlin in 1983, 
the cusp states have been considered to be the antithesis of the Wigner 
crystal and to be described accurately by liquid-like wave functions, such
as the Jastrow-Laughlin \cite{lau2,lau3} (JL) and composite-fermion 
\cite{jai2,jai4} (CF) ones. This view, however, has been recently challenged
\cite{yl1} by the explicit derivation  of trial wave functions for the cusp 
states that are associated with a rotating Wigner (or electron) molecule, RWM. 
As we discussed \cite{yl1} earlier, the RWM wave functions, which are by 
construction {\it crystalline\/} in character, promise to provide a 
simpler, but yet improved and more consistent description of the properties of 
the cusp states, in particular for high angular momenta $(L)$ [corresponding 
to low fractional fillings, since $\nu=N(N-1)/(2L)$].

Issues pertaining to the liquid or crystalline character of the cusp states are
significant in both the fields of QD's and the FQHE. Since the many-body wave 
functions in the lowest Landau level (high $B$) obtained from exact 
diagonalization (EXD), the RWM, and the CF/JL have good angular momenta 
$L=L_0=N(N-1)/2$ \cite{haw}, their electron densities are {\it circularly\/} 
symmetric. Therefore investigation of the crystalline or liquid character of
these states requires examination of the conditional probability distributions
(CPD's, i.e., the fully anisotropic pair correlation functions). These 
calculations were performed here under high magnetic field conditions for QD's 
(in a disk geometry \cite{note3}) with $N=6 - 9$ electrons, 
and for an extensive range of angular momenta. This allowed us to conclude 
that in all instances examined here (corrresponding to  $0.467>  
\nu > 0.111$) the cusp states exhibit an unmistakably crystalline character, 
in contrast to the long held perception in the FQHE literature, with the RWM 
yielding superior agreement with the exact-diagonalization results. 
Furthermore, the RWM states are found to be energetically stabilized 
(i.e., exhibit gain in correlation 
energy) with respect to the corresponding {\it static\/} (symmetry-broken) 
Wigner molecules, from which the multideterminantal RWM 
wavefunctions are obtained through an angular-momentum projection \cite{yl4}.

The CPD's are defined by
\begin{equation}
P({\bf r},{\bf r}_0) =
\langle \Phi | 
\sum_{i \neq j}  \delta({\bf r}_i -{\bf r})
\delta({\bf r}_j-{\bf r}_0) 
| \Phi \rangle  / \langle \Phi | \Phi \rangle,
\label{cpds}
\end{equation} 
where $\Phi ({\bf r}_1, {\bf r}_2, ..., {\bf r}_N)$ 
denotes the many-body wave function under consideration.
$P({\bf r},{\bf r}_0)$ is proportional to the conditional probability of
finding an electron at ${\bf r}$ under the condition that a second electron 
is located at ${\bf r}_0$. This quantity positions the observer on the moving
(intrinsic) frame of reference specified by the collective rotations that are
associated with the good angular momenta of the cusp states.

\begin{figure}[t]
\centering\includegraphics[width=7.cm]{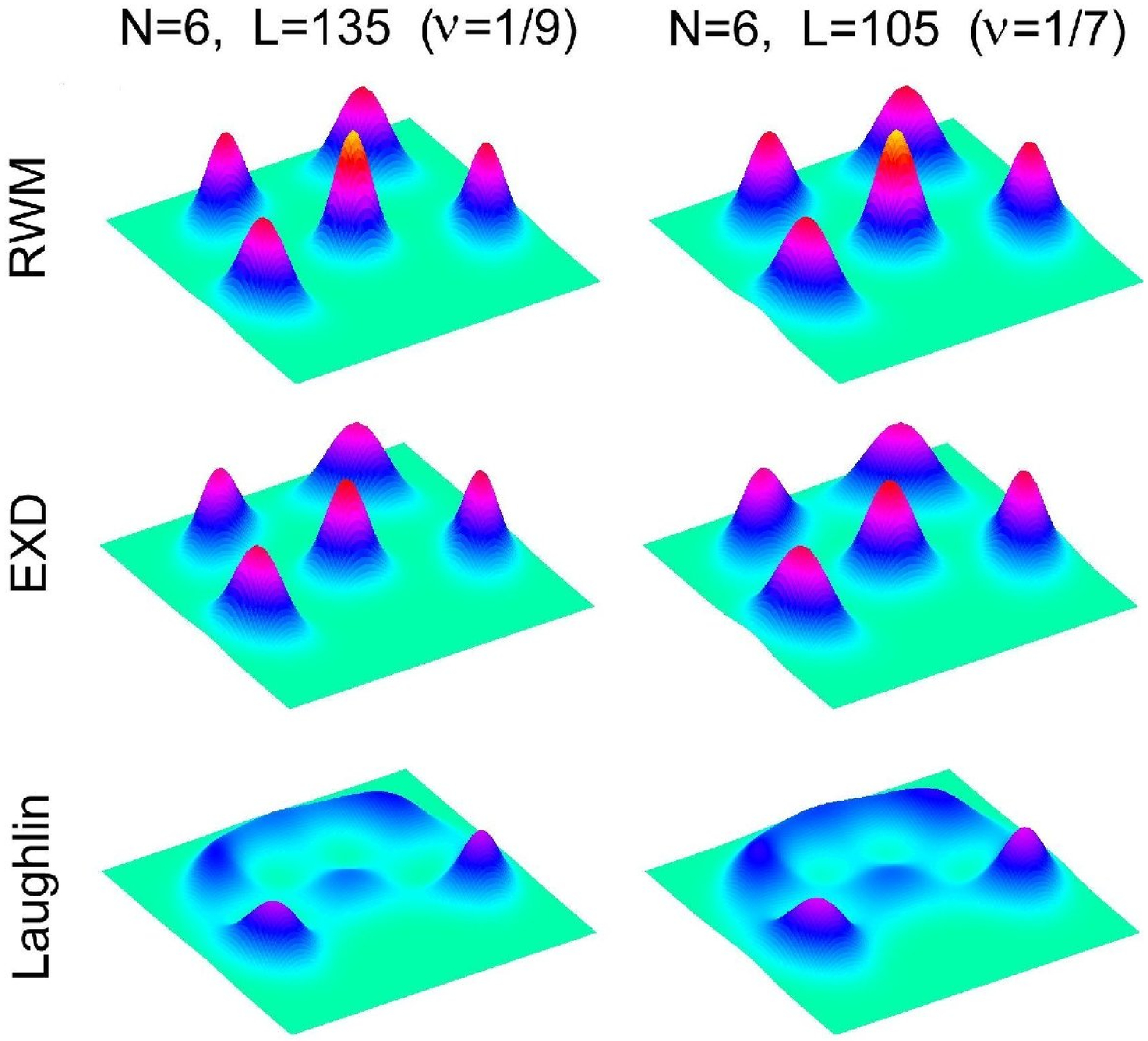}
\caption{
Conditional probability distributions at high $B$ for $N=6$ electrons with 
$L=135$ $(\nu=1/9$, left column) and $L=105$ $(\nu=1/7$, right column). 
Top row: RWM case. Middle row: The case of exact diagonalization. Bottom row:
The Jastrow-Laughlin case.
It is apparent that the exact diagonalization and RWM wave functions have a
pronouned crystalline character, corresponding to the (1,5) polygonal 
configuration of the rotating Wigner molecule. In contrast, the 
Jastrow-Laughlin wave functions fail to capture this crystalline character,
exhibiting a rather "liquid" character. The observation point 
(identified by the missing electron hump) was placed at the 
maximum of the outer ring of the radial electron density \cite{yl1} of the EXD
wave function, namely at $r_0=7.318 l_B$ for $L=135$ and $r_0 = 6.442  l_B$ 
for $L=105$. The EXD Coulomb interaction energies (lowest Landau level)
are 1.6305 and 1.8533 $e^2/\kappa l_B$ for $L=135$ and $L=105$, respectively. 
Here, $l_B=(\hbar c/eB)^{1/2}$. The errors relative to the corresponding EXD 
energies and the overlaps of the trial functions with the EXD ones are: 
(I) For $L=135$, RWM: 0.34\%, 0.860; JL: 0.50\%, 0.665. 
(II) For $L=105$, RWM: 0.48\%, 0.850; JL: 0.46\%, 0.710. 
}
\end{figure}

The CPD's for cusp states corresponding to a lower filling factor than
$\nu = 1/5$, calculated for $N=6$ electrons with $L=135$ ($\nu=1/9$, left 
column) and for $N=6$ with $L=105$ ($\nu=1/7$, right column) are displayed in
Fig.\ 1. Fig.\ 2 displays the CPD's for the cusp states with $N=6$ 
electrons and $L=75$ ($\nu=1/5$, left column) and $N=7$ and $L=105$ 
($\nu=1/5$, right column). In both figures, the top row depicts the RWM case. 
The EXD case is given by the middle row, while the CF cases (which reduce to 
the JL wave functions for these fractions) are given by the bottom row. 

There are three principal conclusions that can be drawn from an inspection
of Figs.\ 1 and 2: 

(I) The character of the EXD states is unmistakably crystalline with the EXD 
CPD's exhibiting a well developed molecular polygonal configuration [(1,5) for
$N=6$ and (1,6) for $N=7$, with one electron at the center], in agreement with 
the explicitly crystalline RWM case. 

\begin{figure}[b]
\centering\includegraphics[width=7.cm]{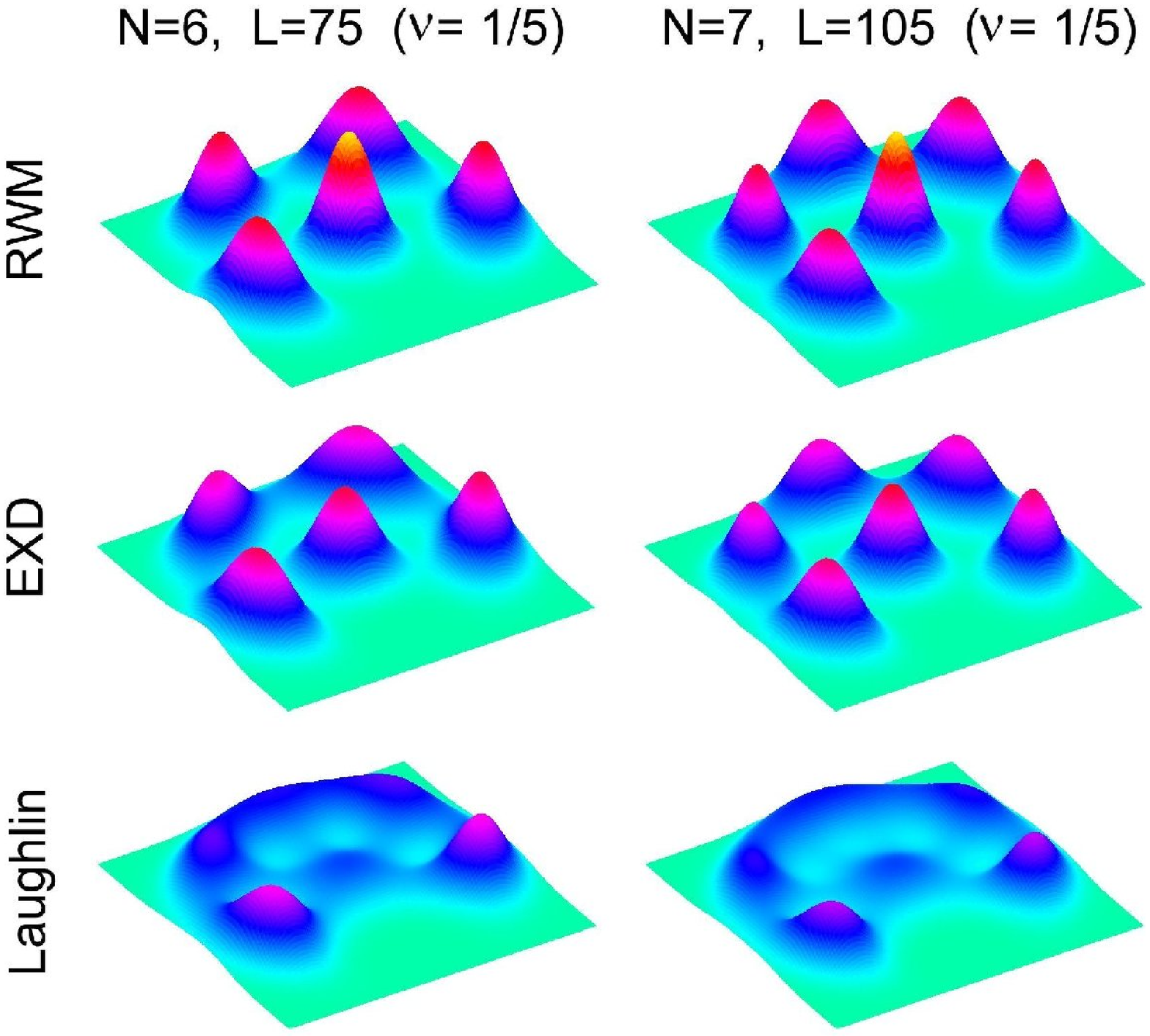}
\caption{
Conditional probability distributions at high $B$ for $N=6$ electrons and 
$L=75$ ($\nu=1/5$, left column) and for $N=7$ electrons and $L=105$ (again 
$\nu=1/5$, right column). Top row: RWM case. Middle row: The case of exact 
diagonalization. Bottom row: The Jastrow-Laughlin case.
The exact diagonalization and RWM wave functions have a pronouned crystalline
character, corresponding to the (1,5) polygonal configuration of the RWM for 
$N=6$, and to the (1,6) polygonal configuration for $N=7$. In contrast, the 
Jastrow-Laughlin wave functions exhibit a characteristic liquid profile that 
depends smoothly on the number $N$ of electrons.
The observation point is located at $r_0 = 5.431 l_B$ 
for $N=6$ and $L=75$ and $r_0=5.883 l_B$ for $N=7$ and $L=105$.
The EXD Coulomb interaction energies (lowest Landau level)
are 2.2018 and 2.9144 $e^2/\kappa l_B$ for $N=6,L=75$
and $N=7,L=105$, respectively. The errors relative to the corresponding EXD
energies and the overlaps of the trial functions with the EXD ones are:
(I) For $N=6,L=75$, RWM: 0.85\%, 0.817; JL: 0.32\%, 0.837. 
(II) For $N=7,L=105$, RWM: 0.59\%, 0.842; JL: 0.55\%, 0.754. 
}
\end{figure}

\begin{figure}[t]
\centering\includegraphics[width=7.cm]{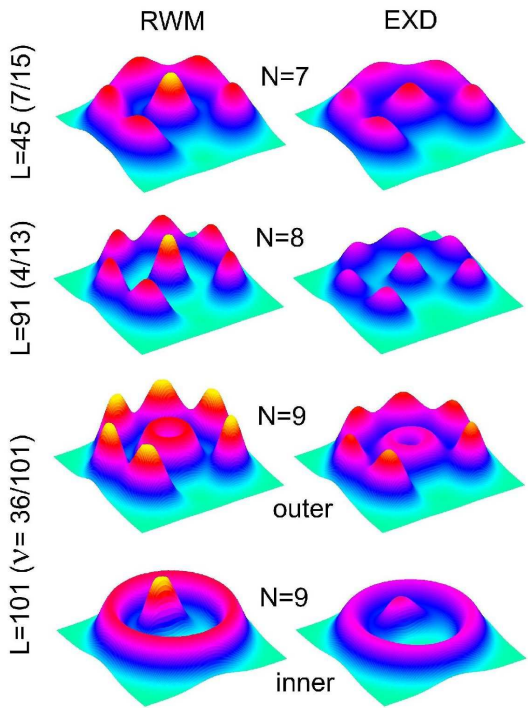}
\caption{
Additional CPD's at high $B$. RWM results: Left column. Results from exact 
diagonalization are depicted on the right column. 
Top row: $N=7$ electrons and $L=45$ ($1/3 < \nu=7/15=0.467 < 1$)
Middle row: $N=8$ electrons and $L=91$ 
($1/5 < \nu=4/13=0.308 < 1/3$).
Two bottom rows: $N=9$ electrons and $L=101$ 
($1/3 < \nu=36/101=0.356 < 1$, see text for explanation).
Even for these low magic angular momenta (high fractional fillings), 
both the exact-diagonalization and RWM wave functions have a pronouned 
crystalline character [corresponding to the (1,6), (1,7), and (2,7) polygonal 
configuration of the RWM for $N=7$, 8, and 9 electrons].
The observation point is located at $r_0=3.776 l_B$ for $N=7,L=45$,
$r_0=5.105 l_B$ for $N=8,L=91$, and $r_0=5.218 l_B$ (outer) and 
$r_0=1.662 l_B$ (inner) for $N=9,L=101$. 
}
\end{figure}

(II) For all the examined instances (covering the 
fractional fillings 1/9, 1/7, and 1/5), the JL wave functions fail to capture 
the intrinsic crystallinity of the EXD states. In contrast, they represent
``liquid'' states in agreement with an analysis that goes back to 
the original papers \cite{lau2,lau3} by Laughlin. In particular, Ref.\ 
\cite{lau3} investigated the character of the JL states through the use of 
a pair correlation function [usually denoted by $g(R)$] that determines the 
probability of finding another electron at the absolute relative distance 
$R=|{\bf r} - {\bf r}_0|$ from the observation point ${\bf r}_0$. 
Our anisotropic CPD of Eq.\ (\ref{cpds}) is of course more 
general (and more difficult to calculate) than the $g(R)$ function
of Ref.\ \cite{lau3}. However, both our $P({\bf r},{\bf r}_0)$ (for $N=6$ and
$N=7$ electrons) and the $g(R)$ (for $N=1000$ electrons, and for $\nu=1/3$ and
$\nu=1/5$) in Ref.\ \cite{lau3} reveal a similar characteristic liquid-like 
short-range-order behavior for the JL states, eloquently described in 
Ref.\ \cite{lau3} (see p. 249 and p. 251). Indeed, we remark that only the 
first-neighbor electrons on the outer rings can be distinguished as separate 
localized electrons in our CPD plots of the JL functions [see Fig.\ 1 and 
Fig.\ 2]. 

(III) The pronounced crystallinity of the EXD states at the rather high 
$\nu=1/5$ value (see Fig.\ 2) does not support the long held expectation 
\cite{lau2,lau3,lam} (based on extrapolations of the JL wave functions to the 
bulk and comparisons with the static bulk Wigner crystal) that a 
liquid-to-crystal phase transition may take place only at the lower fillings 
with $\nu \leq 1/7$. 

Following the conclusion that the crystalline character of the cusp states in 
QD's is already well developed for fractional fillings with the unexpected high
value of $\nu=1/5$, a natural question arises concerning the presence or 
absence of crystallinity in cusp states corresponding to higher fractional
fillings, i.e., states with $ 1/5 < \nu < 1/3$, and even with $ 1/3 < \nu < 1$.
To answer this question, we show in Fig.\ 3 the CPD's for the RWM
(left column) and EXD (right column) wave functions for the case of $N=7$ and 
$L=45$ ($1/3 < \nu=7/15 < 1$), of $N=8$ and $L=91$ ($1/5 < \nu=4/13 < 1/3$),
and of $N=9$ and $L=101$ ( $1/3 < \nu=36/101 < 1$). Unlike the long held 
perceptions in the FQHE literature (which were reasserted in a recent 
manuscript \cite{jai3}), the CPD's in Fig.\ 3 demonstrate that the character 
of the cusp states with high fractional fillings is not necessarily 
``liquid-like''. Instead, these high-$\nu$ cusp states can exhibit a well 
developed crystallinity associated with the (1,6), (1,7), and (2,7) polygonal 
configurations of the RWM, appropriate for $N=7$, 8, and 9 electrons,
respectively. Similar results were also found for the case of $N=6$ electrons.
Note that the case of $N=9$ represents the smallest number of electrons with
a non-trivial concentric-ring arrangement, i.e., the inner ring has more than 
one electrons. As the two CPD's [reflecting the choice of taking the 
observation point (${\bf r}_0$ in Eq.\ ({\ref{cpds})) on the outer or the 
inner ring] for $N=9$ reveal, the polygonal electron rings rotate 
independently of each other. Thus, e.g., to an observer located on the inner
ring, the outer ring will appear as uniform.

Our two-step method for deriving the RWM wave functions is anchored in 
the distinction \cite{yl2} between a {\it static\/} and a {\it rotating} Wigner
molecule, with the collective rotation stabilizing the latter relative to the 
former. Further elaboration on this point requires generation of global ground 
states out of the cusp states, achieved through inclusion \cite{yang,yl2} of an
external parabolic confinement (of frequency $\omega_0$).
In the two-step method, the static WM is first described by an unrestricted 
Hartree-Fock (UHF) determinant that violates the circular symmetry \cite{yl3}. 
Subsequently, the collective rotation of the WM is described by a  
post-Hartree-Fock step of restoration of the broken circular symmetry via 
projection techniques \cite{yl4}. We note that, in the limit 
$N \rightarrow \infty$, the static WM of the UHF develops to the bulk Wigner 
crystal \cite{yosh} and its more sophisticated variants \cite{lam}. 

In general, the localized broken symmetry orbitals of the HF determinant are 
determined numerically via a selfconsistent solution of the UHF equations 
\cite{yl3}. Since we focus here on the case of high $B$, we can approximate 
the UHF orbitals (first step of our procedure) by (parameter free) displaced 
Gaussian functions; namely, for an electron localized at $Z_j$, we use
the orbital
\begin{equation}
u(z,Z_j) = \frac{1}{\sqrt{\pi} \lambda}
\exp \left( -\frac{|z-Z_j|^2}{2\lambda^2} - i\varphi(z,Z_j;B) \right),
\label{uhfo}
\end{equation}
with $z=x+iy$, $Z_j = X_j +i Y_j$, and $\lambda = \sqrt{\hbar /m^* \Omega}$;
$\Omega=\sqrt{\omega_0^2+\omega_c^2/4}$, where $\omega_c=eB/(m^*c)$ is the
cyclotron frequency. The phase guarantees gauge invariance in the presence of
a perpendicular magnetic field and is given in the symmetric gauge by
$\varphi(z,Z_j;B) = (x Y_j - y X_j)/2 l_B^2$, with $l_B = \sqrt{\hbar c/ e B}$.
We only consider the case of fully polarized electrons, which is appropriate at
high $B$.

We take the $Z_j$'s to coincide with the equilibrium positions (forming nested
regular polygons) of $N$ classical point
charges inside an external parabolic confinement of
frequency $\omega_0$, and proceed to construct the UHF determinant
$\Psi^{\text{UHF}} [z]$ out of the orbitals $u(z_i,Z_i)$'s, $i = 1,...,N$.
Correlated many-body states with good total angular momenta $L$ can be
extracted \cite{yl4} from the UHF determinant using projection operators.
We stress that while the initial trial wave function of the UHF equations
consists of a single determinant, the projected wave function is a linear
superposition of many determinants. The projected energy, corresponding to the
symmetry-restored state with angular momentum $L$, is given by 
\cite{yl4}
\begin{equation}
E_{\text{PRJ}} (L) = \left. { \int_0^{2\pi} h(\gamma) e^{i \gamma L}
d\gamma } \right/ { \int_0^{2\pi} n(\gamma) e^{i \gamma L} d\gamma},
\label{eproj}
\end{equation}
with $h(\gamma) =
\langle \Psi^{\text{UHF}}(0) | H | \Psi^{\text{UHF}}(\gamma) \rangle$
and
$n(\gamma) =
\langle \Psi^{\text{UHF}}(0) | \Psi^{\text{UHF}}(\gamma) \rangle,$
where $\Psi^{\text{UHF}}(\gamma)$ is the original UHF determinant rotated by an
azimuthal angle $\gamma$ and $H$ is the many body Hamiltonian (comprised 
\cite{yl4} of the kinetic energy with a vector potential for the magnetic 
field, an external confinement, the Coulomb interelectron repulsion, and the 
Zeeman term). The UHF energies are simply given by 
$E_{\text{UHF}} = h(0)/n(0)$.

\begin{figure}[t]
\centering\includegraphics[width=7. cm]{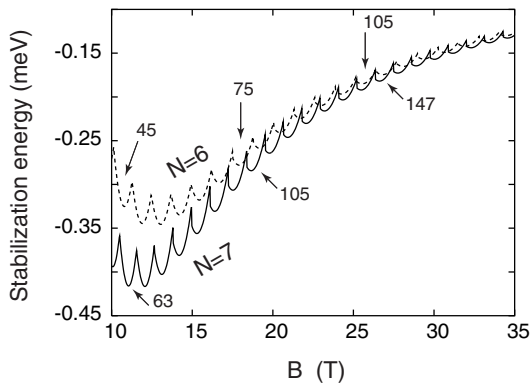}
\caption{Stabilization energies $\Delta E^{\text{gain}}_{\text{gs}}=
E^{\text{gs}}_{\text{PRJ}} - E_{\text{UHF}}$ for $N=6$ (dashed curve) and 
$N=7$ (solid curve) fully polarized electrons in a parabolic QD as a function 
of $B$. The troughs associated with the major fractional fillings (1/3, 1/5, 
and 1/7) and the corresponding ground-state angular momenta are indicated with
arrows. We have extended the calculations up to $B=120$ T (not shown), and 
verified that $\Delta E^{\text{gain}}_{\text{gs}}$ remains negative while its 
absolute value vanishes as $B \rightarrow \infty$.
The choice of parameters is: $\hbar \omega_0=3$ meV (parabolic confinment), 
$m^*=0.067 m_e$ (electron effective mass), and $\kappa =12.9$ (dielectric
constant).
}
\end{figure}

We note that, unlike the UHF ground state (describing a static Wigner
molecule) which has no good angular
momentum, the ground states of the RWM exhibit good angular momenta (labeled
as $L_{\text{gs}}$) that coincide with magic ones [we denote the 
ground-state energy of the RWM as 
$E^{\text{gs}}_{\text{PRJ}} \equiv E_{\text{PRJ}}(L_{\text{gs}})$].

The stabilization of the {\it rotating\/} WM relative to the {\it static\/} 
one [namely the fact that $E^{\text{gs}}_{\text{PRJ}} < E_{\text{UHF}}$,
see Fig.\ 4] is a purely quantum effect. This energy gain
demonstrated here for $N=6$ and 7 electrons, is in fact a general property
of states projected out of trial functions with broken symmetry. This is due
to a theorem \cite{low} stating that at least one
of the projected states (i.e., the ground state) has an energy lower than
that of the original broken-symmetry trial function (e.g., the UHF determinant
considered above).

While our focus here is on the behavior of trial and exact wave functions
in (finite) QD's in high magnetic fields, it is natural to inquire about
the implications of our findings to FQHE systems in the thermodynamic
limit. We recall that appropriate trial wave functions for clean FQHE systems
must possess a good angular momentum $L \geq L_0$, a property shared by
both the CF/JL and RWM functions. We also recall the previous finding that for 
small fractional fillings $\nu$ the static Wigner crystal (which is a 
broken-symmetry state with no good angular momentum) is energetically favored 
\cite{lau2,lau3,jai4,lam} compared to the (liquid-like) CF/JL wave function. 
In contrast, the RWM wave functions remain lower in energy than the 
corresponding static crystalline state for {\it all values\/} of $N$ and $\nu$,
even in the thermodynamic limit. This is due to the fact that the 
aforementioned ``energy-gain'' theorem \cite{low} applies for any number of 
electrons $N$ and for all values of the magnetic field $B$ \cite{note}. 

Since the rotating Wigner crystal carries a current, while the static crystal 
is insulating, we may conjecture that a transition at lower fractional 
fillings from a FQHE conducting state to an insulating Wigner crystal cannot
occur {\it spontaneously\/} for clean systems. Therefore, it should be possible
to observe (see, e.g., Ref.\ \cite{pan}) FQHE behavior at all fractional
fillings in a clean system. In practice, however, impurities and defects, may
pin the RWM, so that one of the main challenges for FQHE observation at such 
low fillings relates to fabrication of high mobility samples \cite{wes,note2}. 

In summary, we have carried out the first systematic investigations
(for $6 \leq N \leq 9$) of cusp states in parabolic quantum dots at high 
magnetic fields. Our anisotropic conditional probability distributions from 
exact diagonalization show that these states are crystalline in character for 
both low and high fractional fillings, unlike the liquid-like Jastrow-Laughlin 
\cite{lau2,lau3} wave functions, but in remarkable agreement with the recently 
proposed rotating-Wigner-molecule \cite{yl1} ones. The cusp states of 
finite-$N$ parabolic QD's are precursors to the bulk fractional quantum Hall 
states (and not to the bulk Wigner crystal) due to the collective rotation, 
which stabilizes the {\it rotating\/} Wigner molecule (having a good angular
momentum) relative to the {\it static\/} one (that exhibits broken symmetry). 

We thank M. Pustilnik for comments on the manuscript.
This research is supported by the U.S. D.O.E. (Grant No. FG05-86ER45234).


\begin{thebibliography}{99}
\bibitem{note3}
The results in Figs.\ 1-3 were obtained without consideration of 
a confinement, and they are not influenced (high $B$) by inclusion 
of a parabolic confinement \cite{jai1,yang,yl2}. For the influence of the
confinement on the energies, see Refs.\ \cite{jai1,yang,yl2}.
\bibitem{note}
We remark that, at present, computational difficulties prohibit
a straightforward extrapolation of energy to the thermodynamic 
limit as a function of $N$ and $\nu$. This is the case especially for the disk
geometry used here, which is advantageous for studies of crystalline
structures, compared to the computationally easier spherical geometry.
\bibitem{note2}
We remark, however, that the stabilization energy diminishes as
$B \rightarrow \infty$ (see Fig.\ 4), and as a result the impurities become
more efficient in pinning the WM for the lower fractional fillings (i.e.,
higher angular momenta).
\end{thebibliography}
\end{document}